\documentclass[preprint]{aastex}
\begin{document}

\title{CCD Photometry of the Globular Cluster NGC 4833 \\ 
  and Extinction Near the Galactic Plane.}

\author{Jason Melbourne and Ata Sarajedini}
\affil{Astronomy Department, Wesleyan University, Middletown, CT 06459}
\email{jmel@astro.wesleyan.edu and ata@astro.wesleyan.edu}

\author{Andrew Layden}
\affil{Dept. of Physics and Astronomy, Bowling Green State University, 
  Bowling Green, OH 43403}
\email{layden@baade.bgsu.edu}
\and
\author{Donald H. Martins\altaffilmark{1}}
\altaffiltext{1}{Observations obtained at the Cerro Tololo Inter-American
Observatory (CTIO), one of the National Optical Astronomy Observatories
(NOAO), which are operated by the Association of Universities for
Research in Astronomy, Inc., under cooperative agreement with the
National Science Foundation.}
\affil{Dept. of Physics, Chemistry and Astronomy, University of
  Alaska, Anchorage, AK 99508}
\email{afdhm@uaa.alaska.edu}

\begin{abstract}
We present CCD photometry for the Galactic globular cluster
NGC 4833. Our BVI color magnitude diagrams 
(CMD) extend from above the red giant branch (RGB) tip to several 
magnitudes below the main sequence turnoff.  The principal sequences 
of the cluster show the 
effects of differential reddening. We have
created a local extinction map, consistent with IRAS and
COBE/DIRBE dust maps of the region.  We use our map to correct the colors
and magnitudes of each star to the value at the cluster center. 
The cluster horizontal branch 
(HB) is predominately blueward of the instability strip with 13 confirmed
RR Lyrae variables and 5 additional RR Lyrae candidates.
Using the 11 confirmed RR Lyraes measured on our images, and the
differential reddening corrected photometry
we calculate $V(HB) = 15.56 \pm 0.063$. 
We have used the simultaneous reddening and metallicity method 
of Sarajedini (1994) to find the mean reddening of the
cluster \(E(B-V) = 0.32 \pm 0.03\), and the mean metallicity 
\([Fe/H] = -1.83 \pm 0.14\). As for the age of NGC 4833, we find 
provisional evidence 
that NGC 4833 is 2 $\pm$ 1 Gyr 
older than the average of M92 and M5.
A more definitive conclusion must await higher resolution imaging of
the cluster core where the effects of differential reddening are minimized.
\end{abstract}


\section{Introduction}
   The Galactic globular cluster NGC 4833 (\(\alpha_{2000}=12^h 59'
34.98" , \,\delta_{2000}=-70^0 52' 28.6"\)) is located in the 
constellation of Musca inhabiting a dusty
region near the Galactic plane \((\ell=304\degr,\,b=-8\degr)\).
Figure 1 (Kohle \& Credner 2000) is an image of the sky showing the features surrounding NGC 4833.
It is clear from this figure that NGC 4833 is likely to suffer from
spatially variable extinction. In fact, based on Washington-band
photoelectric photometry
and high-dispersion spectroscopy of cluster red giants, 
Geisler et al. (1992) and Minniti et al. (1993), respectively, have suggested 
that the reddening varies significantly across
NGC 4833.  This may be one reason for the dearth of CCD photometry 
for this cluster. 

Samus et al. (1995) present the most recent color magnitude
diagram (CMD) of NGC 4833.
Their photographic study includes BV photometry of 868
stars extending from the tip of the red giant branch (RGB) to
the main sequence turnoff and reveals a predominantly blue horizontal
branch (HB).  Accounting for
their measurement errors, the dispersion apparent in the principle
sequences of their CMD is likely to be a result 
of differential reddening.  They find \(V(HB) = 15.6 \pm 0.1\) and a
turnoff magnitude of \(V(TO) = 19.2\).  Menzies (1972) and Alcaino
(1971) give the only other CMDs of NGC 4833.  Menzies
derives a reddening of \(E(B-V) = 0.32 \pm 0.05\) and $V(HB) = 15.5 
\pm 0.05$. The blue HB morphology of NGC 4833 suggests a metal poor cluster 
consistent with Geisler's (1995) value for the mean metallicity 
of \([Fe/H] = -1.84 \pm 0.04\).
Demers \& Wehlau (1977) identify 24 cluster variable stars 
including confirmation of 13 RR Lyraes.  

The CCD photometry presented in Section 3 extends \(\sim 3\)
magnitudes fainter than any previous study, allowing a first
look at the main sequence turnoff of NGC 4833.  The data include 
photometry of the crowded central regions of the cluster unattainable 
in the prior photographic studies.
In addition, we improve on the previous photometry 
by correcting the problematic 
reddening variations across the face of NGC 4833 (Sec. 4). As a result, we are 
able to make more precise measurements of cluster characteristics,
such as the horizontal branch magnitude, reddening, metallicity 
and the age (Sec. 5). An important by-product of our effort is the
production of a reddening map in the region of NGC 4833,
which we can use to investigate the correlation between dust and extinction
using observations from the IRAS and COBE/DIRBE satellites. We begin with a
description of the observations and data reduction.   

\section{Observations and Data Reduction} 

\subsection{Observations}

The data were obtained during two separate observing runs at the Cerro
Tololo Inter-American Observatory (CTIO).  On the night of 1996 May 30,
using the 0.9m telescope and the Tek \#3 2048$\times$2048 pixel CCD, we
obtained fifteen images of NGC 4833 in the B (5 x 120s), V (5 x 60s),
and I (5 x 90s) filters.  The image scale was 0.396 arcsec pixel$^{-1}$ with
a field of view of 13.5 $\times$ 13.5 arcmin.  The telescope was ``dithered'' 
between exposures to minimize systematic errors from chip
defects and flat field features.  Figure 2 shows a V band image of the
cluster. Standard star fields, drawn from the Landolt (1992) list, 
were also observed.
    
Deeper images of NGC 4833 were secured with the CTIO 4.0-m
telescope, PFCCD camera, and the Tek \#4 2048$\times$2048 pixel CCD on
the night of 1995 June 29.  The scale was 0.43 arcsec pixel$^{-1}$.
Three frame pairs were obtained, each consisting of a 200 sec exposure
in $V$ and a 160 sec exposure in $I$.  Two short-exposure frame pairs
were also obtained (10 sec in $V$ and 8 sec in $I$).  In addition the 
pointing of
the telescope was ``dithered'' a few arcsec between each frame pair.
Tests showed that timing corrections for the iris shutter were
unnecessary for images with these exposure lengths.

\subsection{0.9m Data Reduction}

For the photometric reduction of the 0.9m observations, 
we use a process similar 
to that employed by Neely et al.\space(2000); a brief
summary follows. We made use of the DAOPHOT II (Stetson 1994)
crowded-field photometry package. After initial runs of FIND and
PHOT, we construct a preliminary point-spread
function (PSF) from \(\sim\) 40 uncrowded stars.  We use the
preliminary point spread function to subtract away the neighbors
of the full set of \(\sim\)150 PSF stars.  From these, we construct a final PSF
which we use with ALLSTAR to perform profile fitting photometry for all 
the stars on the image. The PSF photometry measures most of the flux
from a star.  We apply an aperture correction to account for any missing
light.  Doing aperture 
photometry on \(\sim\)150 bright \((V \leq 14.5)\) uncrowded stars, 
we calculate 
an aperture correction for each image, looking for spatial variations
over the field.  All fields had a constant correction except for one
of the V images which had a correction that varied linearly with radius
from the center of the image with an amplitude of $\pm$0.01 mag.  
After editing the photometry using the image diagnostics output by
DAOPHOT (Sarajedini \& Da Costa 1991), we combine the aperture 
corrected photometry for stars found in at least 3  of 5 frames taken 
in a given filter. Using stars common to all three filters, 
we compile the BVI photometry list
keeping only those stars with errors that do not exceed 0.1 mag.  The
final calibrated 0.9m photometry (see section 2.4) is presented in
Table 1.

\subsection{4m Data Reduction}
   The 4.0-m images were processed using standard techniques, including
flattening with twilight sky flats.  The FWHM near the center of the
images was typically $\sim$1.0 arcsec, but the image quality degraded
radially from the center.  We therefore masked off the regions of the
chip outside $\sim$6 arcmin from the center.  Photometry of stars
within this radius was obtained using a version of the {\sc DoPHOT}
PSF-fitting photometry code of Schechter et al. (1993) employing a
2-dimensionally variable point spread function 
(Mateo, private communication).  For each
frame pair, we then matched the instrumental photometry of spatially
coincident stellar images.

\subsection{Standard Stars}

The reduction of the 0.9m standard stars and the presentation of
the resultant transformation coefficients have already been discussed
by Montegriffo et al. (1998). We used selected stars from the 
0.9-m $VI$ photometry 
as secondary standards to transform the 4.0-m data to the standard Johnson
$V$ and Kron-Cousins $I$ magnitudes.  To determine the nature of the
transformation required, we plotted the difference between our
instrumental magnitudes ($m$) and the standard magnitude ($M$) as a
function of magnitude, color, and spatial coordinates ($X$ and $Y$).
A weak, linear residual in color and a residual that varied
quadratically in $X$ and $Y$ were present, so we performed a
photometric transformation of the form

\[ m - M = c_0 + c_1 (V-I) + c_2 X + c_3 Y + c_4 X^2 + c_5 XY + c_6Y^2, \]

\noindent where the coefficients $c_i$ were determined by the method of least
squares using over 800 stars spread across the frame.
The root-mean-square of the residuals was 0.03 to 0.05 mag.
The corrections produced by the spatial terms were never more than
$\pm$0.08 mag and were less than $\pm$0.03 mag within 3 arcmin of
the cluster center.  A simpler correction using only linear terms 
in $X$ and $Y$ would not have improved the "U"-shaped pattern of the
residuals. One can think of the spatial terms as a quadratically
varying aperture correction resulting from the radially varying image
quality and the use of the spatially varying PSF.  Photometry from the
individual frame pairs was combined using an error-weighted average.
The results are presented in Table 2.

\section{Color Magnitude Diagrams}

\subsection{0.9m Photometry}

From the data in Table 1, we construct $B-V$ and $V-I$ 
CMDs as shown in Fig. 3. Based on the width of the principle
cluster sequences, the proximity of the cluster to the Galactic plane, 
and the appearance of IRAS  and COBE/DIRBE maps of NGC 4833 we find
strong evidence for differential reddening across the 
frame.  In addition, non-cluster stars contaminate the CMD 
with a characteristic
column occurring at \((B-V) \sim 1.0\).  
Figure 4 gives the CMD for stars within 3 arcmin of the cluster
center.  The predominately blue HB morphology in conjunction with
the relatively steep RGB indicate a low cluster metal abundance. 

\subsection{4m Photometry}

Figure 5 shows the deep $V-I$ CMD yielded by the 4m observations
in Table 2. Two artifacts of our reduction procedure are seen in
Fig. 5. First, there is an overabundance of bright stars in the 
CMD due to the fact that the fainter stars are significantly incomplete
near the center of the cluster. Second, the splice between the short
and long exposures appears as a discontinuity in the CMD at $I=17.0$ mag.
Taken together, these irregularities simply mean that we cannot
reliably use the cluster luminosity function, but the locations of
principal features in the CMD are unaffected. 

Our photometry reveals the main sequence (MS) extending to V $\sim$23.5, or 
$\sim$3 magnitudes deeper than Samus et al. (1995).
Figure 5 also illustrates the CMD for stars between 1.5 and 3 arcmin 
of the cluster center, where we find the MS turnoff at V $\sim$19.

\subsection{Comparison to Previous Photometry}

With 528 stars in common, we compare our 0.9m photometry to that of
Samus et al. (1995) in Fig. 6.  We find a mean V offset of \(0.192 \pm
0.007\), and a mean $B-V$ difference of \(\Delta(B-V)= -0.026 \pm 0.003\).
In addition, there are clear nonlinear distortions present in Fig. 6.
We have seen similar behavior in previous papers
(Sarajedini \& Norris 1994; Sarajedini \& Layden 1995; Neely et al. 2000) 
when we compare CCD and photographic photometry.

\section{Differential Reddening Map}    

\subsection{Generating the Map}

In order to correct for the  differential reddening across the frame, we
need a two-dimensional reddening map of the cluster.  
To produce such a map, we began by fitting a fiducial sequence to the blue
HB.  In the case of little or no differential
reddening, HB stars will cluster tightly about the fiducial, spread out only
by the photometric errors and evolution away from the zero age horizontal
branch.  
The 0.9m field was divided into square sections and a CMD was constructed
for each section as shown in Fig. 7a and 7b. To ensure an adequate
number of blue HB stars in each section, the squares around the edge of the 
frame are larger, \(3.3 \times 3.3\) arcmin, than the interior squares,
which are \(2.6 \times 2.6 \) arcmin. Nonetheless, there is one square
that contains no blue HB stars.
For each square, we calculate the mean $\Delta(B-V)$ necessary to 
bring each blue HB star in the 0.9m photometry
onto the fiducial under the constraint that
$\Delta$$V$ = 3.2 $\Delta(B-V)$.\footnote{Note that we have 
adopted the traditional value of $R_V = 3.2$. This
agrees to within the errors with other
determinations of $R_V$, which suggest a value closer to 3.1 
(Cardelli et al. 1989). The results of our analysis do not depend
sensitively on the precise value of $R_V$.}  This color correction is
then associated with the (X,Y) position of the center of each square.
We use stepwise regression (Kleinbaum et al. 1988) to fit the data 
with a polynomial of the following form 
\begin{eqnarray}
\Delta(B-V) = c_0 + c_1X + c_2 X^2 + c_3 Y + c_4 Y^2 + c_5XY,
\end{eqnarray}
where the (X,Y) positions are given in pixels.
The coefficients of the fit, which features a root-mean-square
deviation of 0.019 mag, along with their associated errors are
given in Table 3. Note that the values of $c_2$ and $c_5$ were not 
statistically significant and have been dropped from the fit.

\subsection{Comparison to IRAS and COBE/DIRBE Maps}

Figure 8 shows our local reddening and extinction maps
compared to dust maps generated by the COBE/DIRBE
\footnote{COBE datasets were developed by the NASA Goddard Space Flight Center 
under the guidance of the COBE Science Working Group and were provided by 
the NSSDC.} satellite at 100 $\mu$m and the 
IRAS\footnote{NASA IPAC/Jet Propulsion Laboratory} satellite at
60 $\mu$m as downloaded from
SkyView\footnote{SkyView was developed and is maintained under NASA 
ADP Grant NAS5-32068 with P.I. Thomas A. McGlynn under the auspices of the High Energy Astrophysics Science Archive Research Center (HEASARC)
at the GSFC Laboratory for High Energy Astrophysics.}.
While the large scale distribution of dust (Fig. 1)
suggests a higher reddening toward the
Galactic plane (northeast), our map (Fig. 8a and 8b)
and the dust maps produced by IRAS and COBE/DIRBE (Fig. 8c and 8d)
suggest that the small scale reddening gradient is in the exact opposite
direction.  The reddening in our 13.5 arcmin field increases away from 
the Galactic plane (southwest).  The lower panels of Fig. 8 show the result
of dividing our extinction map by the IRAS (Fig. 8c) and
COBE/DIRBE (Fig. 8d) dust maps scaled to unit mean. The latter maps
are fairly coarse which means that the extinction pattern present in
our map appears as a distinct remnant in the resultant quotient.
However, we find remarkably good agreement in the overall shape between 
our photometrically derived extinction map
and the aforementioned dust maps. The typical deviation is approximately
$\pm$10\%.  The agreement is maximized in the center of
the frame and then deviates near the edges. 
 
\subsection{Applying the Reddening Map}

Our strategy in applying the reddening map to our photometry involves
the application of a color and magnitude correction calculated using
Equation 1 combined with $\Delta$$V$ = 3.2 $\Delta(B-V)$ and
$\Delta(V-I) = 1.345 \Delta(B-V)$ (Dean et al. 1978). This correction brings
the photometry of each star in the field to the value it would have
if it was located at the center of the cluster.
Figures 9, 10 and 11 show radial CMD's of our 0.9m and 4m data
before and after the differential reddening correction. 
All of the principal sequences are significantly better defined after
the differential reddening correction.  
Because these are radial CMDs, we see the maximum improvement in the
radial annuli that are furthest from the cluster center. In all of
the subsequent analysis, we will make exclusive use of this differential
reddening corrected photometry.

\section{Cluster Properties}

\subsection{The Horizontal Branch}

Demers \& Wehlau (1977) investigated the nature of 24 variable stars in 
NGC 4833. Based on an analysis of their light curves,
they established or confirmed the presence of 13 RR Lyrae variables. We have
photometry for 11 of these stars (Table 4), shown 
as circles in Fig. 12.  All 11 appear on the HB.  
These stars give a mean HB magnitude of $V(HB) = 15.56 \pm 0.063$, 
in good agreement with the Samus et al. (1995) value $V(HB)=15.6 \pm 0.1$.
However, we caution that our V(HB) value is based on only a few
observations that do not properly sample the light curves of the
RR Lyrae stars.  A proper CCD-based time-series analysis is required for a 
definitive value of V(HB).

We have photometry for 7 of the remaining 11 variable stars studied by 
Demers \& Wehlau (1977). One of these was listed as a `red
variable' by Demers \& Wehlau and the remaining 6 were of unknown
type.  These additional variable stars are plotted as triangles in
figure 12.  We confirm
that star \#16, listed as a red variable, is near the tip of the RGB.
Star \#9, unclassified by Demers, is also near the tip of the RGB and may
be another red variable.  We find that the remaining five
unclassified stars (\#'s 3, 7, 12, 15, and 24) are located on the  
HB in the instability strip and are most likely RR Lyrae variables.
When we include the five additional RR Lyraes, we find a  
$V(HB)= 15.56 \pm 0.059$, which is unchanged from our previous estimate.  

Because NGC 4833 possesses a predominantly blue HB and a significant
population of RR Lyrae variables,
it provides an opportunity to test the fiducial HB technique used by
Sarajedini (1994a), Montegriffo et al. (1998) and Neely et al.
(2000). The technique is used to estimate V(HB) for a cluster with a
negligible RR Lyrae population but with a significant blue HB. It relies
on the fact that the downward curvature of the blue HB is mainly a result
of the increasing V band bolometric corrections at bluer colors. 
Fiducial blue HBs of well-observed clusters with RR Lyrae populations,
are fit to the blue HB of the cluster in
question. The expected location of the RR Lyraes is then inferred from
such a comparison. Using a technique that is identical to that of
Neely et al. (2000), we performed such a fit to the blue HB of NGC 4833
using fiducials from the globular clusters M15 (Buonanno, Corsi, \& Fusi
Pecci 1985) and M79 (Ferraro et al. 1992) in $B-V$ and 
M92 and M5 (Johnson \& Bolte 1998) in $V-I$.
The weighted average gives $V(HB) = 15.60 \pm 0.056$, which supports
our value based on the cluster RR Lyraes.

\subsection{Reddening and Metallicity}

We have applied the simultaneous reddening and metallicity (SRM) method
of Sarajedini (1994b; Sarajedini \& Layden 1997). The SRM method is based 
on two empirically derived
relationships. The first relates metallicity to the dereddened color of
the RGB at V(HB) and the second correlates metallicity to the difference
in magnitude between V(HB) and the RGB at a dereddened color of 1.2.  
Along with a polynomial describing the shape and location of the RGB,
these relations allow an internally consistent measurement of the
reddening and metallicity. Taking the average of the SRM results in
$B-V$ and $V-I$, we derive a 
cluster metallicity of $[Fe/H] = -1.83 \pm 0.14$, on the Zinn \& West
(1984) scale, and a mean reddening of
$E(B-V) = 0.32 \pm 0.03$. These results compare favorably with 
those of Menzies (1971) who obtained $E(B-V) = 0.32 \pm 0.05$ from a
UBV two-color diagram and with the extinction maps of 
Schlegel et al. (1998), which yield $E(B-V) = 0.33$ at the position
of NGC 4833. As for previously measured metal abundance values,
Geisler et al. (1995) and Rutledge et al. (1997) both used 
Calcium triplet spectroscopy of giant stars and found 
$[Fe/H] = -1.84 \pm 0.04$ and $[Fe/H] = -1.92 \pm 0.02$, respectively;
both of these estimates are in good agreement with our metallicity.

\subsection{Cluster Age}

\subsubsection{Isochrone Comparisons}

To investigate the age of NGC 4833, we will follow two approaches.
First, there is the classical isochrone fitting technique. We make
use of the Girardi et al. (2000) models along with the RR Lyrae distance
scale advocated by Chaboyer (1999):
\begin{eqnarray}
M_V(RR) = (0.23 \pm 0.04) ([Fe/H] + 1.6) + (0.56 \pm 0.12).
\end{eqnarray}
In order to make the isochrone comparisons as differential as possible,
we will also perform the same technique on the globular clusters M5
and M92. Table 5 contains the relevant quantities for the clusters in
question. The values for M92 and M5 are taken from Layden \& Sarajedini
(2000, and references therein). Figure 13 shows the isochrone fits
to M5 and M92 while Fig. 14 displays the fit to NGC 4833, where we
have utilized the differential reddening corrected photometry for
stars located between 100 arcsec and 180 arcsec of the cluster center. For all
three clusters, the unevolved main sequence is fit
reasonably well by the isochrones, but the isochrones are systematically
too red as compared with the data along the lower giant branch.
This is not surprising given the fact that the mixing length parameter,
which controls the efficiency of convective transport and thus the
colors of these convectively unstable stars, is still a major
uncertainty in the stellar models (Chaboyer et al. 1996). Examining
the fit primarily in the turnoff region suggests that M5 has an age of
$13 \pm 1$ Gyr, M92 is $14 \pm 1$ Gyr old, and NGC 4833 is 
$15 \pm 2$ Gyr old. The estimated uncertainties are purely internal
errors based only on the scatter of points in the turnoff region
and do not include possible uncertainties in the models or errors
in the adopted distance moduli, reddenings, or metallicities. 
The age errors are substantial making it difficult to reach a definitive
conclusion. However, in the next section, we investigate a more
quantitative age determination technique.

\subsubsection{The Magnitude of the Subgiant Branch}

In Layden \& Sarajedini (2000), we used an age determination technique
that relies upon the magnitude difference between the HB and the 
subgiant branch ($\Delta$$V_{SGB}^{HB}$) at a point 0.21 mag 
bluer than $(V-I)_g$. This is
primarily useful when the turnoff region is difficult to define because
of field contamination. Originally, this age diagnostic was applied to
the CMD of the globular cluster M54, which suffers from contamination
from field stars belonging to the Sagittarius dwarf galaxy and bulge
stars in the line of sight. In the
present context, it is useful to apply it to NGC 4833 because
of the heavy contamination from Galactic disk stars in the cluster's 
CMD. 

Using the NGC 4833 photometry shown in Fig. 14, we construct a luminosity
function centered at $(V-I) = (V-I)_g - 0.21 = 1.15$ with a width of
$\pm0.03$ mag. Fitting a Gaussian to this distribution yields
$V(SGB) = 18.44 \pm 0.02$ which then gives 
$\Delta$$V_{SGB}^{HB}$=$2.88 \pm 0.07$.
Using the RR Lyrae distance scale described above, we follow
the technique described by Layden \& Sarajedini (2000) in order to
parameterize age as a function $\Delta$$V_{SGB}^{HB}$ and metallicity. 
To effect this calibration, we have utilized two different grids of
models - the Revised Yale Isochrones (Green et al. 1987, RYI) to allow a
comparison to the age scale of Layden \& Sarajedini (2000) and the
most recent Girardi et al. (2000, GirI) isochrones representing the `state
of the art.' Our results are shown in Table 5. We note that both the
RYI and GirI suggest that NGC 4833 is $2 \pm 1$ Gyr older than M92 and
M5. However, this conclusion is rather tentative until higher resolution
imaging of the cluster core becomes available allowing the 
construction of a CMD that does not require such dramatic corrections
for differential reddening.

\section{Conclusions}
   
We presented CCD photometry for the Galactic globular cluster
NGC 4833.  The cluster is located behind a
dusty region southwest of the Galactic plane.  Interestingly, the 
local reddening increases to the southwest, 
away from the Galactic plane. Our reddening map is in good agreement
with dust emission as measured by the IRAS and COBE/DIRBE satellites
with a maximum deviation of $\pm$10\%.
After correcting our photometry for the effects of differential
reddening we used the simultaneous reddening and metallicity method 
of Sarajedini (1994) to find the mean reddening of the
cluster \(E(B-V) = 0.32 \pm 0.03\) and the mean metallicity 
\([Fe/H] = -1.83 \pm 0.14\) on the Zinn \& West (1984) scale.  
The cluster contains 13 RR Lyrae
variables with confirmed periods.  Using the 11 RR Lyraes available in
our photometry, we calculated $V(HB) = 15.56 \pm 0.063$.  Five
additional unclassified variables appear on the HB in the instability
strip and are probably RR Lyraes.  
Finally, we used our main sequence photometry to determine an age for 
NGC 4833 relative to those of M92 and M5. We found provisional evidence 
that NGC 4833 is 2 $\pm$ 1 Gyr
 older than the comparison clusters.
A more definitive conclusion must await higher resolution imaging of
the cluster core where the effects of differential reddening are minimized.

\break

\figcaption{Photograph of the Crux/Musca/Carina region taken from
the compendium of constellations obtained by Kohle \& Credner (2000).
Several deep sky objects are indicated. Note that NGC 4833 lies
in an especially dusty region near the plane of the Milky Way.}
 
\figcaption{Image of the globular cluster NGC 4833 in the V band
taken with the CTIO 0.9m telescope.  The field of view is
13.5 $\times$ 13.5 arcmin. North is up and East is to the left.}

\figcaption{Color-magnitude diagrams of the entire field observed
around NGC 4833 from the 0.9m data. The left panel shows the
$B-V$ data while the right panel shows $V-I$.} 
 
\figcaption{Same as Fig. 3 except that only stars within 3 arcmin of the
  cluster center are plotted.}
 
 \figcaption{Color-magnitude diagrams of the  field observed
around NGC 4833 from the 4m observations. The entire frame is plotted in the
   left panel while only stars between 1.5 and 3.0 arcmin are plotted in the 
right panel.}
 
\figcaption{Comparison  of our 0.9m photometry with that of 
Samus et. al. (1995).  The top panel shows the $V$ mag offset, while the bottom 
plots the $(B-V)$ color offset.}
 
\figcaption{Our observed 0.9m field, represented by the outer box, is 
divided into square subsections.  For each square, a $B-V$ CMD is
  plotted.  The size and location of each
  square corresponds to its area in the full frame. Panel (a) shows the size 
and
  location of the outer squares while (b) plots the inner squares.  The
  inner squares can be smaller because there are more stars in the CMD.  
North is up and East is to the left.  The solid line represents our
adopted blue HB fiducial.}
 
\figcaption{(a) The derived reddening map contours given in units of $E(B-V)$. 
(b) The derived extinction map contours given in $V$
magnitudes. (c) Contours of the IRAS 60 $\mu$m  image in the direction
of NGC 4833. The
contours are listed in units of MegaJanskys/steradian.
(d)  The COBE/DIRBE image in the direction of NGC 4833.  The 13.5$\times$13.5 
arcmin field is covered by 4 pixels.  The pixel values are given in units
of MegaJanskys/steradian. 
  (e) The IRAS $60\mu$m  image of NGC 4833 divided by our extinction map
and scaled  to unit mean.  (f) The COBE/DIRBE image of NGC
  4833 divided by our extinction map and scaled to unit mean.  
  The residual of our extinction map appears in this figure because 
  of the coarse resolution in the COBE/DIRBE image.}
 
\figcaption{Radial $B-V$ CMD's of NGC 4833 from the 0.9m data.  
  The left panel shows the cluster CMD
  before correcting for differential reddening.  The right shows the
  CMD after the correction.}
 
\figcaption{Same as Fig. 9 except that the $V-I$ CMDs are plotted.}
 
\figcaption{Same as Fig. 10 except that the 4m data are plotted.}
 
\figcaption{Variable stars in the CMDs of NGC 4833.  The circles are
  confirmed RR Lyraes, while the triangles are additional variable stars
  listed in Demers \& Wehlau (1977).  The left panel shows the $B-V$
CMD, and the right shows $V-I$.}

\figcaption{The Girardi et al. (2000) isochrones fit to the comparison
  clusters M5 (left) and M92 (right). The distances and reddenings used are
shown in Table 5. Ages of $13 \pm 1$ Gyr and $14 \pm 1$ Gyr are
inferred for M5 and M92, respectively. }
 
\figcaption{The Girardi et al. (2000) isochrones fit to NGC 4833. The adopted
distance and reddening are given in Table 5.  An Age of $15 \pm 2$ Gyr 
is inferred for NGC 4833.}

\begin{deluxetable}{c c c c c c}
\tablewidth{3.5in}
\tablecaption{NGC 4833 0.9m Photometry}
\tablehead{
 \colhead{star} &   \colhead{X}  & \colhead{Y} & \colhead{V} &
 \colhead{B--V}&\colhead{V--I}}
\tablecolumns{6}
\startdata
1 &  702.09 & 2042.79 & 16.16 & 1.24 & 1.38\\
2 &  781.77 & 2042.20 & 14.97 & 1.15 & 1.29\\
3 &  530.44 & 2041.66 & 18.64 & 0.58 & 0.86\\
4 &  725.23 & 2040.56 & 18.58 & 0.91 & 1.09\\
5 & 1559.94 & 2039.18 & 16.85 & 0.65 & 0.91\\
6 & 1071.77 & 2038.73 & 16.91 & 0.80 & 0.98\\
7 &  834.37 & 2038.64 & 17.01 & 0.86 & 1.02\\
8 &  347.12 & 2038.15 & 18.17 & 0.97 & 1.24\\
9 & 1775.14 & 2037.27 & 17.40 & 0.88 & 1.05\\
10 & 1276.10 & 2036.62 & 17.76 & 0.80 & 1.10\\
\enddata
\tablenotetext{1}{Note. - Table 1 is presented in its entirety in the
  electronic edition of the Astronomical Journal.  A portion is
  presented here for guidance regarding its form and content.}
\end{deluxetable}

\begin{deluxetable}{c c c c c}
\tablewidth{3in}
\tablecaption{NGC 4833 4m Photometry}
\tablehead{
 \colhead{star} &   \colhead{X}  & \colhead{Y} & \colhead{V} &
 \colhead{V--I}}
\tablecolumns{5}
\startdata
1 &  888.03 & 1864.86 & 18.25 & 1.02 \\
2 &  818.92 & 1840.18 & 16.39 & 1.01 \\
3 &  723.30 & 1823.80 & 18.24 & 1.03 \\
4 & 1048.15 & 1819.82 & 17.65 & 1.30 \\
5 &  992.68 & 1818.87 & 22.05 & 1.12 \\
6 & 1078.56 & 1817.79 & 14.64 & 1.29 \\
7 &  783.26 & 1817.35 & 16.41 & 1.39 \\
8 & 1219.86 & 1816.27 & 21.93 & 1.29 \\
9 & 1188.82 & 1816.21 & 21.42 & 1.21 \\
10 & 1176.73 & 1815.56 & 22.89 & 1.53\\
\enddata
\tablenotetext{1}{Note. - Table 2 is presented in its entirety in the
  electronic edition of the Astronomical Journal.  A portion is
  presented here for guidance regarding its form and content.}
\end{deluxetable}

\begin{deluxetable}{c c c }
\tablewidth{2.5in}
\tablecaption{Reddening Surface Coefficients}
\tablehead{
 \colhead{Variable} &   \colhead{Coefficient}  & \colhead{Error} }
\tablecolumns{4}
\startdata
  $c_0$ & 0.0067 & 0.0042  \\
  $c_1$ & -0.0450 & 0.0084\\
  $c_2$ & 0.0405 & 0.0079\\
  $c_3$ & 0 & 0\\
  $c_4$ & 0.049 & 0.017\\
  $c_5$ & 0 & 0\\
\enddata
\end{deluxetable} 
\begin{deluxetable}{c c c }
\tablewidth{5in}
\tablecaption{Variable Stars in NGC 4833}
\tablecolumns{2}
\tablehead{
 \colhead{Star ID\#  } &   \colhead{Star ID\# in
   Demers }
  & \colhead{Type of Variable }\\
\colhead{in Table 1  } &   \colhead{(Demers et al. 1977) }
  & \colhead{}}
\tablecolumns{3}
\startdata
1566   & 3 &    Unconfirmed RR Lyrae\\
1432   & 4 &    RR Lyrae \\
2237   & 5 &    RR Lyrae \\
1655   & 7 &    Unconfirmed RR Lyrae\\
1740   & 9 &    Unconfirmed Red Variable\\
1583   & 12 &   Unconfirmed RR Lyrae\\
2036   & 13 &   RR Lyrae \\
2140   & 14 &   RR Lyrae \\
2294   & 15 &   Unconfirmed RR Lyrae\\
651   & 16 &   Red Variable \\
2673   & 17 &   RR Lyrae \\
2677   & 18 &   RR Lyrae\\
2331   & 19 &   RR Lyrae\\
1676   & 20 &   RR Lyrae\\
810   & 21 &   RR Lyrae\\
2498   & 22 &   RR Lyrae\\
1152   & 23 &   RR Lyrae\\
2057  & 24 &   Unconfirmed RR Lyrae\\
\enddata
\end{deluxetable}

\begin{deluxetable}{c c c c c c c}
\tablewidth{6.5in}
\tablecaption{Cluster Parameters}
\tablehead{
 \colhead{Cluster} &   \colhead{[Fe/H]} & \colhead{V(HB)}  & \colhead{E(V--I)} &
\colhead{$\Delta$V} & \colhead{Age(RYI)} & \colhead{Age(GirI)} }
\tablecolumns{4}
\startdata
 M92      & $-2.24 \pm 0.08$ & $15.10 \pm 0.03$ & 0.03 & $2.79 \pm 0.04$ & $14.5 \pm 0.5$ & $13.8 \pm 0.6$ \\
 M5       & $-1.40 \pm 0.06$ & $15.09 \pm 0.02$ & 0.04 & $2.78 \pm 0.04$ & $14.5 \pm 0.7$ & $13.4 \pm 0.6$ \\
 NGC 4833 & $-1.83 \pm 0.14$ & $15.56 \pm 0.06$ & 0.43 & $2.88 \pm 0.07$ & $16.6 \pm 1.2$ & $15.5 \pm 1.0$ \\
\enddata
\end{deluxetable} 

\end{document}